\documentclass{article}
\usepackage{spconf,amsmath,graphicx,hyperref}
\usepackage{booktabs}
\usepackage{multirow}
\usepackage{xcolor}
\usepackage{subcaption}

\title{Assessing the Impact of Speaker Identity in Speech Spoofing Detection}
\twoauthors
 {Anh-Tuan DAO, Driss Matrouf }
	{Laboratoire d’informatique d’Avignon, France}
 {Nicholas Evans}
	{EURECOM, Sophia Antipolis, France}

\begin{document}
%
\maketitle
\begin{abstract}
Spoofing detection systems are typically trained using diverse recordings from multiple speakers, often assuming that the resulting embeddings are independent of speaker identity. However, this assumption remains unverified. In this paper, we investigate the impact of speaker information on spoofing detection systems. We propose two approaches within our \emph{Speaker-Invariant Multi-Task} framework, one that models speaker identity within the embeddings and another that removes it. SInMT integrates multi-task learning for joint speaker recognition and spoofing detection, incorporating a gradient reversal layer.
Evaluated using four datasets, our speaker-invariant model reduces the average equal error rate by 17\% compared to the baseline, with up to 48\% reduction for the most challenging attacks (e.g., A11).


\end{abstract}
\begin{keywords}
audio deepfake detection, speaker-aware, speaker-invariant, SSL model
\end{keywords}
\section{Introduction}
\label{sec:intro}

Automatic Speaker Verification (ASV) systems offer secure biometric recognition based on voice characteristics, but sophisticated spoofing techniques mimicking legitimate users pose significant security challenges. To counter these threats, spoofing detection systems are increasingly deployed to protect ASV integrity. Prior work has explored multi-task learning (MTL) to enhance spoofing detection, with ResNet-based models demonstrating improved performance for ASVspoof 2019 by incorporating speaker recognition as an auxiliary task~\cite{MTL-SSD-MoW22}. These results indicate that speaker-related information provides complementary cues that help discriminate between bona fide and spoofed utterances. 

MTL approaches combining speaker recognition and spoofing detection have been more commonly applied in spoofing-aware speaker verification (SASV) tasks~\cite{SASV2022}, but attract less attention in standalone spoofing detection. 
One reason could be the limited number of speakers in spoofed datasets like ASVspoof 2019~\cite{ASVspoof19} which, despite providing speaker labels, lacks sufficient diversity for robust speaker-related training. Prior work~\cite{MTL-SASV2023} addressed this through multi-stage training: first pre-training models using large-scale speaker recognition datasets like VoxCeleb2, followed by fine-tuning using the ASVspoof 2019 dataset with multi-task objectives. The recent ASVspoof 5~\cite{ASVspoof24} database, with its expanded set of speaker IDs, supports direct MTL for spoofing detection, motivating our investigation into unified training paradigms optimized jointly for both tasks.


Recent self-supervised learning (SSL) models significantly outperform traditional architectures by leveraging large-scale pre-training to capture rich speech representations~\cite{AASIST,conformer,MHFA_Spoof,rohdin24_but_asvspoof}.
Spoofing detection systems are generally trained on varied recordings from multiple speakers, typically assuming that generated embeddings are independent of speaker identity. This hypothesis is questionable and is dependent on the training data, particularly when the number of speakers or their diversity is limited. 

In this paper, we investigate the role of speaker identity in spoofing detection systems. Specifically, our goal is to determine whether explicitly incorporating speaker-related information enhances performance or, conversely, degrades it. Additionally, we explore whether suppressing speaker-specific information could be beneficial. To address these questions, we propose the \textbf{{Speaker-Invariant Multi-Task}} (SInMT) framework, an SSL-based architecture which exploits SSL-based feature extraction with Multi-Head Factorized Attention (MHFA~\cite{MHFA_ASV}) classifiers. SInMT supports flexible switching between two configurations. The first, \textbf{Speaker-Invariant (MHFA-IVspk)}, employs a gradient reversal layer (GRL) with two classification heads, one for binary spoof detection and one for speaker identification (using speaker IDs), to adversarially minimize speaker discriminability. By suppressing speaker-specific information, the model focuses on spoofing indicators. The second, \textbf{Speaker-Aware (MHFA-spk)}, jointly trains spoofing detection and speaker identification tasks without GRL, leveraging speaker-dependent cues to improve detection in scenarios where spoofs exhibit mismatched speaker characteristics.
The framework enables seamless switching between configurations by activating or deactivating the GRL, allowing direct comparisons of speaker-aware and speaker-invariant strategies within a unified SSL backbone.

Our main contributions are:
\begin{itemize}
\item \textbf{SInMT framework}: We investigate the impact of speaker information on spoofing detection systems by proposing a flexible SSL-based framework. The SInMT unifies speaker-aware and speaker-invariant approaches for spoofing detection, built using SSL representations with dual MHFA classifiers. We will show that handling speaker information, either by removing it or integrating it, improves performance.
\item \textbf{Comprehensive empirical evaluation}: We demonstrate that both configurations consistently outperform the baseline MHFA model, with MHFA-IVspk reducing the average equal error rate (EER) by 17.2\% across four datasets, and delivering substantial improvements for the most challenging attack types (e.g., a 48\% EER reduction for attack A11).

\end{itemize}

\section{SSL-based Spoofing Detection}

Our approach to spoofing detection uses SSL models to extract representations from raw audio. 
The model architecture includes two primary components: an SSL-based feature extractor and a back-end classifier. The feature extraction module comprises a CNN encoder and a Transformer-based contextual network. The CNN encoder transforms raw audio into latent representations $z_{1:T}$, compressing dimensionality. These representations are then processed by the Transformer-based contextual network, yielding context-aware representations $o_{1:T}$. 
The Transformer's self-attention mechanism effectively models long-range dependencies critical for interpreting speech patterns. 
We assess and benchmark three back-end classifiers as baselines: Conformer~\cite{conformer}, MHFA~\cite{MHFA_Spoof}, and AASIST~\cite{wave2vec2-seft-learning}.

\section{Speaker-Invariant Multi-Task Framework for Spoofing Detection}

We propose the SInMT framework for speech spoofing detection, designed to either leverage or suppress speaker information through a flexible architecture. SInMT enables two configurations: a speaker-aware (MHFA-spk) that jointly optimizes spoofing detection and speaker recognition, and a speaker-invariant (MHFA-IVspk) that uses a GRL to suppress speaker-specific patterns.

\subsection{Architecture}

\begin{figure}[t]
\centering
\includegraphics[width=1.0\linewidth]{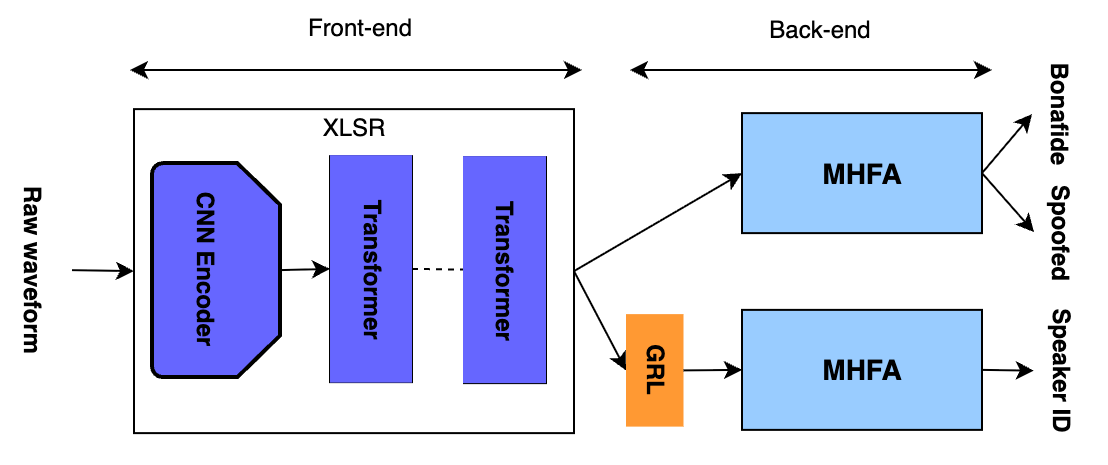}
\caption{Our model architecture with dual classifiers and GRL.}
\label{fig:idfe_arch}
\end{figure}

The SInMT framework uses a SSL backbone and MHFA classifiers, as depicted in Figure~\ref{fig:idfe_arch}. The MHFA classifiers used in the SInMT framework are identical to those of the MHFA baseline model. SInMT comprises:
\begin{itemize}
\item \textbf{Feature Extractor:} A pre‐trained XLSR encoder~\cite{XLS-R2022} that maps raw audio to a sequence of contextualized frame‐level embeddings.
\item \textbf{Spoofing Classifier Head:} A MHFA network that uses the XLSR embeddings to predicts the binary label $\hat{y}_s\in \{\text{bonafide},\text{spoof}\}$.
\item \textbf{Speaker Classifier Head:} A second MHFA network (structurally identical to the spoofing head) that predicts the speaker ID $\hat{y}_d\in{1,\dots,D}$, where $D$ is the number of speakers used during training.
\item \textbf{Gradient Reversal Layer (GRL):} Inserted between the feature extractor and the speaker classifier head. In the forward pass, the GRL acts as an identity. During backpropagation, it multiplies the gradients by a hyperparameter $-\lambda$ to encourage the feature extractor to remove speaker‐specific patterns.
\end{itemize}
For MHFA-spk, the spoofing and speaker heads are trained jointly with a gradient scaling factor \(\lambda=-1\) to leverage speaker-specific cues. 

\label{sec:sinmt}

\subsection{Speaker-Invariant Adversarial Training}

To generate embeddings that are invariant to speaker identity, we employ an adversarial training objective. For an input audio signal $x$, the feature sequence is denoted as $o = f(x)$. The spoofing classifier is represented by $g_s(\cdot)$, and the speaker classifier by $g_d(\cdot)$. The training process optimizes the following objective:

\begin{equation}
\min_{f, g_s, g_d} \mathcal{L}_s(g_s(f(x)), y_s) + \alpha \mathcal{L}_d(g_d(\mathrm{GRL}(f(x))), y_d)
\label{eq:adv_obj}
\end{equation}

Here, the terms are defined as:
\begin{itemize}
    \item $\mathcal{L}_s$: cross-entropy loss for classifying spoofing attempts.
    \item $\mathcal{L}_d$: cross-entropy loss for speaker identification.
    \item $\alpha$: a hyperparameter that adjusts the balance between spoofing detection and speaker invariance.
\end{itemize}




\begin{table*}[h]
\centering
\renewcommand{\arraystretch}{0.5} 
\caption{Model performance comparison in term of Equal Error Rate (EER).}
\label{table:1}
\begin{tabular}{lcccccc}
\toprule
& \multicolumn{5}{c}{EER} \\
\midrule
\textbf{Model} & \textbf{ITW} & \textbf{ASV5 eval} & \textbf{ASV21LA} & \textbf{ASV21DF}  & \textbf{Pooled} \\
\midrule
AASIST & 7.03 & 5.54 & 13.66 & 9.60 & 8.95 \\
Conformer & 5.69 & \textbf{3.85} & 12.49 & 10.40 & 8.10 \\
MHFA & 4.31 & 4.64 & 12.14 & 8.58 & 7.41 \\
MHFA-spk & 3.76 & 5.29 & 8.67 & 8.41 & 6.53 \\
MHFA-IVspk & \textbf{3.58} & 4.98 & \textbf{8.41} & \textbf{7.57} & \textbf{6.13} \\
\bottomrule
\end{tabular}
\end{table*}

\begin{table*}[h]
\centering
\renewcommand{\arraystretch}{0.5} 
\caption{Model performance (\% EER) of different models, evaluated on ASVspoof 2021 LA break down: attack types (A07-19). The \textit{pooled} indicates the average performance across all categories.}
\label{table:2}
\begin{tabular}{lcccccccccccccc}
\toprule
\textbf{Model} & \textbf{A07} & \textbf{A08} & \textbf{A09} & \textbf{A10} & \textbf{A11} & \textbf{A12} & \textbf{A13} & \textbf{A14} & \textbf{A15} & \textbf{A16} & \textbf{A17} & \textbf{A18} & \textbf{A19} & \textbf{Pooled} \\
\midrule
MHFA & 1.54 & 1.91 & 0.76 & 20.77 & 17.02 & 3.45 & 4.75 & 3.82 & 1.49 & 4.32 & 7.01 & 19.67 & 37.57 & 12.14 \\
MHFA-spk & 1.07 & 1.14 & 0.67 & 14.40 & 10.30 & 3.72 & 6.61 & 1.94 & 1.15 & 3.72 & 4.93 & 11.18 & 28.61 & 8.67 \\
MHFA-IVspk & 0.88 & 1.15 & 0.45 & 12.56 & 8.76 & 3.52 & 3.61 & 2.36 & 1.12 & 3.32 & 4.76 & 15.45 & 24.85 & 8.41 \\
\bottomrule
\end{tabular}
\end{table*}

\subsection{Parameter Optimization via Backpropagation}

Model parameters are updated using stochastic gradient descent to achieve speaker-invariant features for spoofing detection. Let $\theta_f$, $\theta_{g_s}$, and $\theta_{g_d}$ denote the parameters of the XLSR feature extractor, spoofing classifier, and speaker classifier, respectively. The updates are performed as follows:

\begin{equation}
\theta_f \leftarrow \theta_f - \mu \left( \frac{\partial \mathcal{L}_s^i}{\partial \theta_f} - \lambda \frac{\partial \mathcal{L}_d^i}{\partial \theta_f} \right)
\label{eq:feat_update}
\end{equation}
\begin{equation}
\theta_{g_s} \leftarrow \theta_{g_s} - \mu \frac{\partial \mathcal{L}_s^i}{\partial \theta_{g_s}}
\label{eq:spoof_update}
\end{equation}
\begin{equation}
\theta_{g_d} \leftarrow \theta_{g_d} - \mu \frac{\partial \mathcal{L}_d^i}{\partial \theta_{g_d}}
\label{eq:speaker_update}
\end{equation}

In Equations (\ref{eq:feat_update}) to (\ref{eq:speaker_update}), $\mu$ represents the learning rate, and $\lambda$ is a hyperparameter that controls the weight of the speaker-invariant loss~\cite{pmlr-2015-GRL}. The feature extractor update in (\ref{eq:feat_update}) minimizes the spoofing classification loss $\mathcal{L}_s^i$ while maximizing the speaker classification loss $\mathcal{L}_d^i$ through the GRL, which scales the speaker classifier gradient by $-\lambda$ during backpropagation. This outs to confuse the speaker classifier, thereby promoting speaker invariance. Equations (\ref{eq:spoof_update}) and (\ref{eq:speaker_update}) update the spoofing and speaker classifiers, respectively, using standard gradient descent to optimize $g_s(\cdot)$ and $g_d(\cdot)$. 

\begin{figure*}[t]
    \centering
    \begin{subfigure}[b]{0.3\linewidth}
        \centering
        \includegraphics[width=0.99\linewidth]{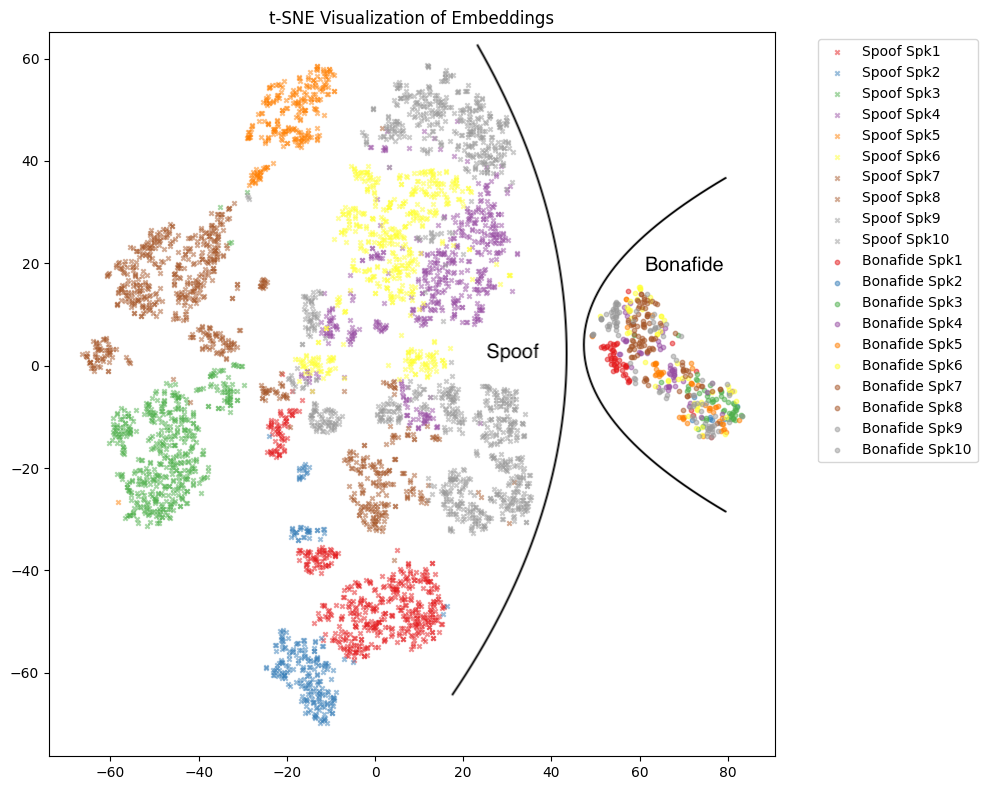}
        \caption{MHFA.}
        \label{fig:learned_emb_a}
    \end{subfigure}
    \hfill
    \begin{subfigure}[b]{0.3\linewidth}
        \centering
        \includegraphics[width=0.99\linewidth]{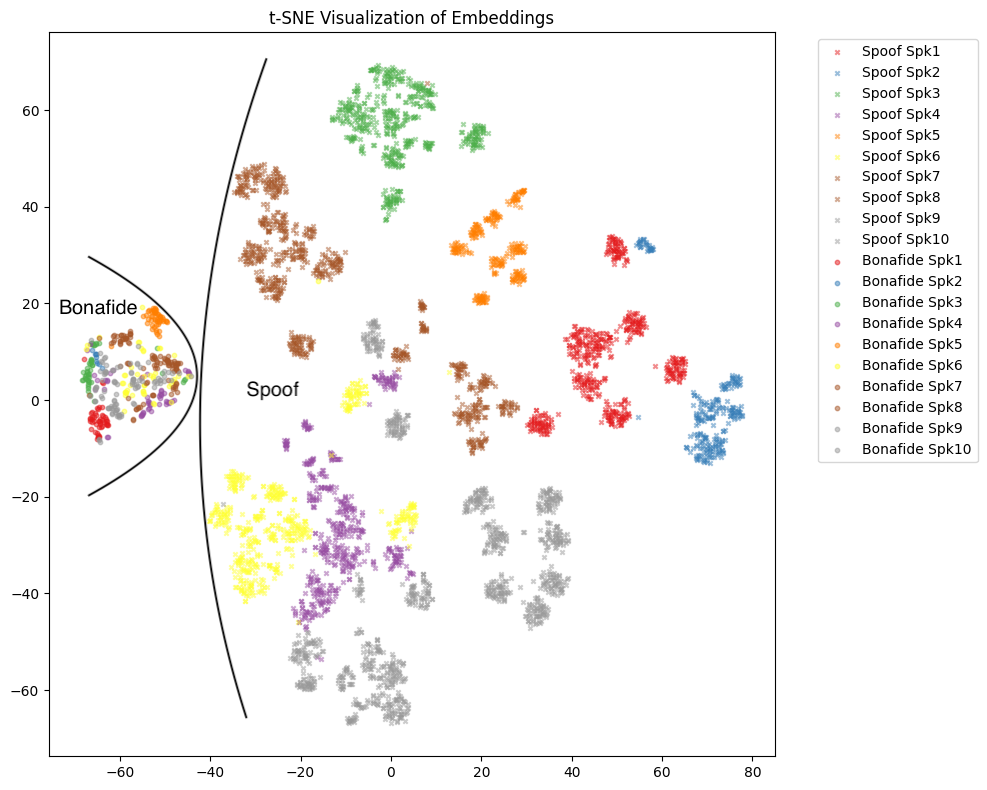}
        \caption{MHFA-spk.}
        \label{fig:learned_emb_b}
    \end{subfigure}
    \hfill
    \begin{subfigure}[b]{0.3\linewidth}
        \centering
        \includegraphics[width=0.99\linewidth]{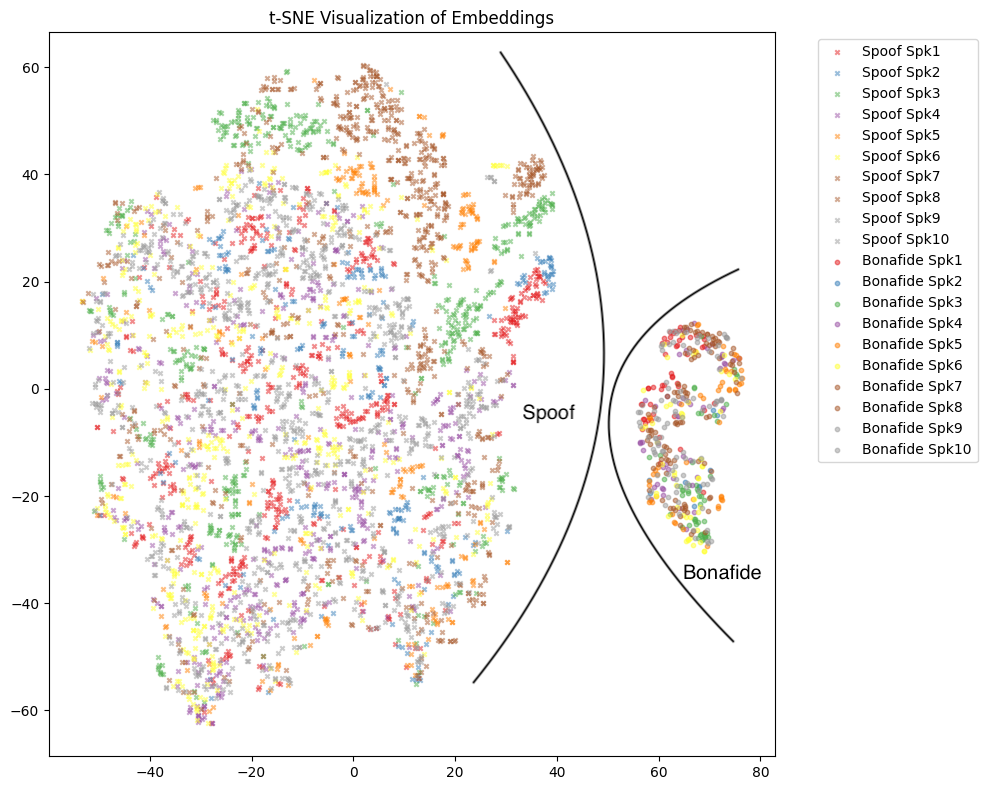}
        \caption{MHFA-IVspk.}
        \label{fig:learned_emb_c}
    \end{subfigure}

    \caption{t-SNE visualization of embeddings from MHFA, MHFA-spk and MHFA-IVspk models, showcasing audio representations of ten distinct speakers from the ASVspoof 5 dataset, each represented by a unique color.}
    \label{fig:learned_emb}
\end{figure*}
\section{Experimental Setup}
\subsection{Datasets and Metrics}\label{sec:datasets-training}
We train models using the ASVspoof 5 training set which comprises approximately 180,000 utterances collected from 400 speakers. For evaluation, we use four datasets: ASVspoof 5 evaluation set, In-the-Wild (ITW)~\cite{Wild2022}, ASVspoof 2021 LA and DF hidden subsets~\cite{ASVspoof21hid, ASVspoof21}. These hidden subsets have non-speech segments removed, preventing models from exploiting such shortcuts. Prior work~\cite{Martin-DonasARG24} shows the challenges presented by these hidden subsets, compared to standard evaluation subsets, since they demand reliance on core spoofing cues rather than non-speech-related shortcuts. Performance is evaluated using the EER.

\subsection{Data Augmentation}
During training, we employed common data augmentation strategies using the MUSAN dataset and the real room impulse response (RIR) database~\cite{MUSAN,Reverb2017}. Each training utterance is enhanced with one of four augmentation techniques: reverberation, speech, music, or noise (\cite{dao24_asvspoof}, Section 3.2). For consistency, all experiments reported in this paper adopt the same data augmentation protocol.


\subsection{Implementation Details}\label{sec:implem_details}
During training, input utterances are randomly segmented into 4-second clips. For evaluation, models operate on complete audio clips. We use the Adam optimizer~\cite{adam} with a learning rate of $10^{-6}$ to minimize a weighted cross-entropy loss. Training is conducted in batches of 32 samples over 30 epochs using NVIDIA A100 GPUs. The hyperparameter $\alpha$ is set to 0.1. We train the MHFA-spk model with a gradient scaling factor \(\lambda=-1\), leveraging speaker-specific cues. Subsequently, the MHFA-IVspk model is initialized from the pre-trained MHFA-spk model, and then trained with the GRL-enabled adversarial objective using \(\lambda=1\) to enforce speaker invariance.

\section{Results}
We compare the performance of three baseline models (AASIST, Conformer, and MHFA), then choose the best (MHFA) for experiments with our SInMT framework.

\subsection{Model Performance Comparision}
Table~\ref{table:1} presents performance for four datasets.
Among the baseline models, MHFA achieves the best average performance, reducing the pooled EER by 17.2\% compared to AASIST. The most significant improvement is observed for the ITW dataset, where MHFA reduces the EER from 7.03\% to 4.31\% (38.7\% relative gain). Consequently, we apply our SInMT framework to the MHFA classifier to further enhance performance.

We now compare MHFA-spk and MHFA-IVspk against the MHFA baseline. MHFA-spk improves the pooled EER by 11.8\% (from 7.41\% to 6.53\%), with the largest gain for the ASVspoof 2021 LA dataset, reducing the EER from 12.14\% to 8.67\%. MHFA-IVspk further improves performance, achieving a 17.2\% reduction in pooled EER compared to MHFA (from 7.41\% to 6.13\%). Notable improvements are seen for out-of-domain datasets: ASVspoof 2021 LA (30.7\% relative gain, 12.14\% to 8.41\%) and ITW (17.0\% relative gain, 4.31\% to 3.58\%). 
MHFA-IVspk slightly outperforms MHFA-spk, but their condition-specific advantages are unclear. We suspect that MHFA-IVspk excels on spoofing datasets with unseen speakers due to its speaker-invariant training, while MHFA-spk may be superior on datasets with seen speakers. As the evaluation datasets feature disjoint speakers, future work will evaluate MHFA-spk on closed-set speaker datasets to confirm these distinctions.

\subsection{Visualisation}
A visualisation of embeddings using t-SNE is shown in Fig.~\ref{fig:learned_emb}.
Specifically, we apply t-SNE on the embeddings of audio samples from 10 distinct speakers selected at random from the ASVspoof 5 training dataset.
In the embedding space for the MHFA model, we observe a partial separation between speaker IDs, though clusters remain closely packed. This suggests some retention of speaker-specific information. For the MHFA-spk model, the visualization shows more pronounced clustering by speaker identity, indicating that this model better preserves speaker-specific information. 
Notably, each speaker ID forms small sub-clusters, potentially reflecting individual speaker characteristics. 
In contrast, MHFA-IVspk, trained with a GRL to suppress speaker discriminability, shows no distinct speaker clusters, corresponding to the objective of minimising speaker-specific information while focusing on spoofing-related information cues.

\subsection{Attack Types Breakdown Analysis}
An analysis of model performance across different attack types in the ASVspoof 2021 LA dataset is presented in Table~\ref{table:2}. The dataset includes diverse attack types, ranging from text-to-speech and voice conversion to hybrid approaches. The attack types exhibit performance variability, reflecting their distinct synthesis characteristics. 
Our proposed models, MHFA-spk and MHFA-IVspk, demonstrate performance gains across all attack types compared to the MHFA baseline. The most substantial improvements are observed for A10 and A11, where the high naturalness and speaker similarity typically pose significant detection challenges. Specifically, compared to MHFA, MHFA-IVspk achieves a 40\% relative EER reduction for A10 (from 20.7\% to 12.5\%) and a 48\% relative EER reduction on A11 (from 17.0\% to 8.7\%).

\section{Conclusion}
This study demonstrates the critical role of speaker-specific information in enhancing spoofing detection systems. We found that handling speaker information, either by removing it or integrating it, is crucial. Our proposed SInMT framework, leveraging the MTL with a GRL, effectively integrates speaker identification and spoofing detection. The SInMT enables flexible switching between speaker-aware (MHFA-spk) and speaker-invariant (MHFA-IVspk) configurations. Assessed across four datasets, both configurations outperform the MHFA baseline, with the MHFA-IVspk model reducing the pooled EER by 17\%. Future work will explore a hybrid approach that dynamically integrates speaker-aware and speaker-invariant models. 


\section{Acknowledgements}
This project was provided with computing HPC and storage resources by GENCI at IDRIS to thank the grant 2025-AD011016488 on the supercomputer Jean Zay's A100 partition. This work was financially supported by ANR BRUEL (ANR-22-CE39-0009).


\bibliographystyle{IEEEbib}
\bibliography{strings,refs}

\end{document}